\documentclass[]{elsart}

\usepackage{epsfig}

\usepackage{amssymb}

\newcommand{\dir}{Figs}

\newcommand{\rr}{ {\bf r} }
\newcommand{\vv}{ {\bf v} }
\newcommand{\nn}{ {\bf n} }
\newcommand{\ff}{ {\bf f} }
\newcommand{\EE}{ {\bf E} }
\newcommand{\erfc}{ {\mbox{erfc}} }

\begin{document}

\begin{frontmatter}




\title{Mechanisms of DNA separation in entropic trap arrays: 
       A Brownian dynamics simulation}

\author{Martin Streek, Friederike Schmid, Thanh Tu Duong, Alexandra Ros}

\address{
Fakult\"at f\"ur Physik, Universit\"at Bielefeld, 
33615 Bielefeld, Germany 
}

\begin{abstract}
Using Brownian dynamics simulations, we study the migration 
of long charged chains in an electrophoretic microchannel 
device consisting of an array of microscopic entropic traps
with alternating deep regions and narrow constrictions.
Such a device has been designed and fabricated recently by 
Han \etal~for the separation of DNA molecules 
(Science, 2000). Our simulation reproduces the
experimental observation that the mobility increases
with the length of the DNA. A detailed data analysis allows
to identify the reasons for this behavior. Two distinct 
mechanisms contribute to slowing down shorter chains. 
One has been described earlier by Han \etal: the chains 
are delayed at the entrance of the constriction and 
escape with a rate that increases with chain length.
The other, actually dominating mechanism is here reported 
for the first time: Some chains diffuse out of their main 
path into the corners of the box, where they remain trapped 
for a long time. The probability that this happens increases 
with the diffusion constant, {\em i.~e.}, the inverse chain
length.
\end{abstract}

\begin{keyword}
gel electrophoresis \sep microfluidic system \sep DNA separation
\sep entropic trap \sep computer simulation
%\PACS 61.30.Hn \sep 61.30.Vx 
\end{keyword}
\end{frontmatter}

\section{Introduction}
\label{sec:introduction}

Electrophoresis is the standard method of DNA separation by 
length (Viovy, 2000). Since the mobility of DNA molecules 
in free solution is independent of their length, the DNA is
commonly introduced into a gel, which serves as a random sieve. 
Unfortunately, the efficiency of gel electrophoresis 
decreases with the length of the DNA. Moreover, bioanalytic 
devices are more and more miniaturized, and incorporating
random gels with characteristic pore sizes in the nanometer 
range into future nanodevices will presumably become 
increasingly problematic. Therefore, much recent effort 
has been devoted to designing and making well-defined 
microstructured devices for DNA separation 
({\em e.~g.}, Han \etal, 1999; 2000; 2002; Bader \etal, 1999; 
 Hammond \etal, 2000; Duong \etal, 2003).

In this paper, we focus on a geometry proposed by Han and
coworkers (Han \etal, 1999), which is based on the idea
of entropic trapping. The DNA is introduced into a small
microchannel with alternating deep regions and shallow 
constrictions. The channel thickness in the constrictions
is much smaller than the radius of gyration of the DNA molecules. 
The deep region is large enough to accommodate the equilibrium
shape of the DNA molecules. Entering the constrictions
is thus entropically penalized, and the deep regions can 
be interpreted as entropic traps.
A schematic cartoon of the structure is shown in 
Fig.~\ref{fig:device}.

\newcommand{\CCdevice}
{
\caption{
Schematic cartoon of the microchannel device proposed by 
Han \etal~(2000) (side view). The width of the channel 
in the $y$-direction is much larger ($\sim 30 \mu m$).
}
\label{fig:device}
\bigskip
}
\begin{figure}[t]
\centerline{\includegraphics[width=0.9\textwidth]{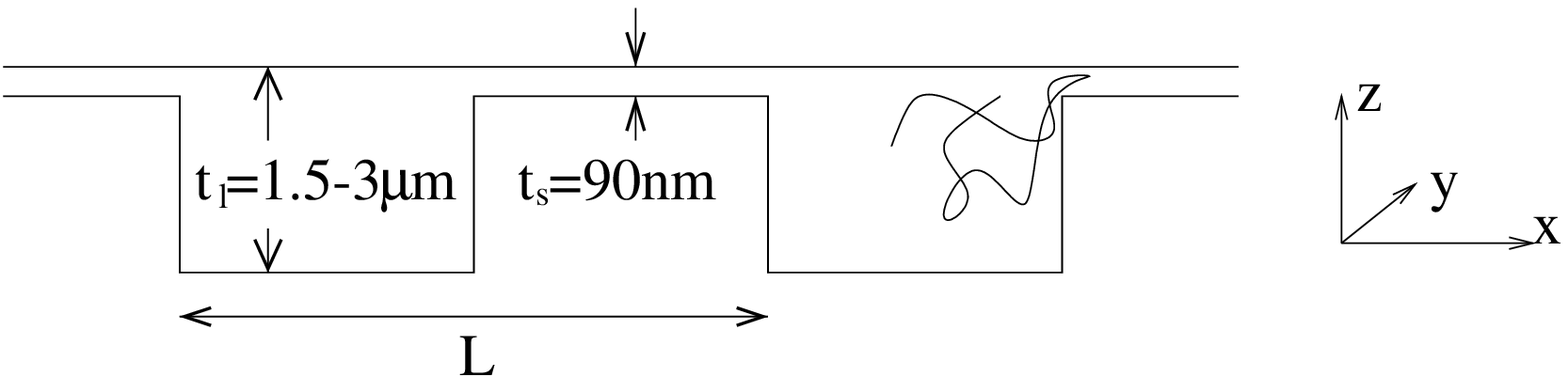}}
\CCdevice
\end{figure}

In these structures, Han \etal~made the counterintuitive 
observation that long DNA molecules travel {\em faster} 
than short molecules. They explained their finding by means 
of a simple kinetic model.
In order to escape from one of the deep regions, 
the DNA must overcome an activation barrier $\Delta F_c$. 
The escape process is initiated by a thermally activated
stretching of a chain portion (length $x$) into the 
shallow constriction. This costs entropic penalty 
proportional to $x$, but it also decreases the electric 
potential energy by an amount $\sim E x^2$, where $E$ is 
the electric field.  Hence there exists a critical length 
$x_c \propto 1/E$, below which the entropy 
penalty dominates and the chain is driven back 
into the deep region. At $x > x_c$, the energy
gain dominates and the chain escapes. The free energy
at $x_c$ represents the activation barrier for the 
escape process. It depends solely on the electric
field, \mbox{$\Delta F_c \propto 1/E$}. 
The rate of escape attempts, $1/\tau_0$, increases
with the chain size, since the amount of polymer in 
contact with the constriction is larger for 
larger chains. The mean trapping time in this
simple model is given by
\begin{equation}
\label{eq:tau}
\tau = \tau_0 \exp(\Delta F_c/k_B T),
\end{equation}
where $T$ is the temperature and $k_B$ the Boltzmann 
factor. This implies that long chains travel faster. 

The results of Han \etal~have motivated recent 
Monte Carlo simulations by Tessier \etal~(2002).
These authors used the Bond fluctuation model, a well-known
lattice model for polymers, to study the mobility
of polymers in a geometry similar to that suggested by
Han \etal~within local Monte Carlo dynamics 
(single monomer moves). Their results confirm the 
trapping picture of Han \etal, and even details 
of the kinetic model. For example, they present evidence 
that the penetration depth $x$ of the chain into the 
constriction can be used as a ``reaction coordinate'' 
for escaping, with a critical value $x_c \propto 1/E$.

When looking more closely at the simulation data, one 
notices that the trapping in the system of Tessier \etal~is 
unexpectedly strong. Although the width of the 
constriction is more than twice as large as in the experiments 
(in units of the persistence length), the molecules spend 
almost the entire time in a trapped state, and very little 
time traveling, even at intermediate fields. This effect
might be an artifact of the lattice model. We note that 
the width of the constriction in the simulations is 
just 10 lattice constants, while every monomer occupies
a cube with 8 lattice sites, and the average bond 
length is 2.8 lattice constants. Moreover, the Monte 
Carlo dynamics is not realistic. Monomers are picked 
and moved randomly (with some moves being rejected), 
whereas in the real system, they are pulled into
the constriction by the electric force. Dynamic Monte
Carlo is known to reproduce diffusional and relaxational 
dynamics in systems near local equilibrium. Nevertheless,
it is not clear how well it can be applied to study 
driven systems. The details of the dynamics should matter 
particularly at higher electric fields, in situations where 
the chains travel so fast that they no longer reach local 
equilibrium in the traps. Therefore, simulations 
of off-lattice models with a more realistic dynamical 
evolution are clearly desirable. 

As a first step in this direction, Chen and Escobedo 
(2003) have recently studied the free energy landscape
of chains in a single, non-periodic trap with Monte 
Carlo simulations of an (off-lattice) bead-spring model.
The initial configuration in these simulations
is that of a fully relaxed chain in the absence
of an electric field. With that starting point,
the free energy barrier $\Delta F$ turns out to
depend on the chain length for short chains,
and to level off at higher chain lengths. The data
for $\Delta F$ do not seem to support the relation 
$\Delta F \propto 1/E$. Since the simulations still
used Monte Carlo, the results on dynamical properties
were limited.

In the present paper, we present Brownian dynamics 
simulations of a full (non-equilibrium) periodic array 
of entropic traps. We employ a Rouse-chain model, 
such as has been successfully used by others
to study the migration of DNA in various geometries
(Deutsch, 1987; 1988; Matsumoto, 1994; Noguchi, 2001).
Our main result is the observation of a new trapping
mechanism in geometries such as Fig.~\ref{fig:device},
which is at least as important as that suggested by
Han \etal, and which can also be exploited to achieve
molecular separation. The insight gained from our study 
should be useful for the design of future, improved 
molecular separation devices.

The paper is organized as follows: in the next
section, we describe and discuss the model and
give some technical details on the simulation. 
The results are presented in section \ref{sec:results}.
We summarize and conclude in section \ref{sec:summary}.
 
\section{The Simulation Model}

\label{sec:model}

We model a single DNA molecule as a chain with pairwise interactions
\begin{equation}
\label{eq:lennard-jones}
V_{pair}(r)/k_B T = \left\{
\begin{array}{lcr}
(\sigma/r)^{12}-(\sigma/r)^6 + 1/4 
&:& (r/\sigma)^6 \le 2 \\
0 &:& \mbox{otherwise}.
\end{array}
\right.
\end{equation}
This potential describes soft, purely repulsive interactions
between beads of diameter $\sigma$. The beads are connected 
by springs with the spring potential
\begin{equation}
V_{spring}(r) = \frac{k}{2} r^2.
\end{equation}
The spring constant was chosen very large, $k = 100 k_B T/\sigma^2$, 
in order to prevent chain crossings. 

The molecule is confined in a structured channel with
a geometry similar to that used by Han \etal.
The thickness of the shallow constrictions and thick regions is 
$t_s = 7 \sigma$ and $t_l = 80 \sigma$, respectively, the
length of a deep region is 80 $\sigma$, and the total length 
of a period is $L=100 \sigma$. In the lateral direction, 
the channel is infinite. The chain beads are repelled
from the walls of the channel by means of a wall potential 
essentially identical to (\ref{eq:lennard-jones}),
\mbox{$V_{Wall}(r) = V_{pair}(r)$}, where $r$ is the distance
of a bead to the closest wall point\footnote{At corners, the
potentials of the adjacent walls are summed up}. 
The details of the wall potential do not influence the 
results, as long as it is repulsive and short ranged.
Note that the effective width of the channel, {\em i.~e.},
the width of the space accessible to a bead, is reduced 
by $2 \sigma$ with this potential. A snapshot of a 
chain in such a channel is shown in Fig. \ref{fig:snapshot}.

\newcommand{\CCsnapshot}
{
\caption{
Snapshot of a chain with $N=1000$ beads in an entropic trap.
The solid lines show the electric field lines.
}
\label{fig:snapshot}
\bigskip
}
\begin{figure}[t]
\centerline{\includegraphics[width=0.7\textwidth]{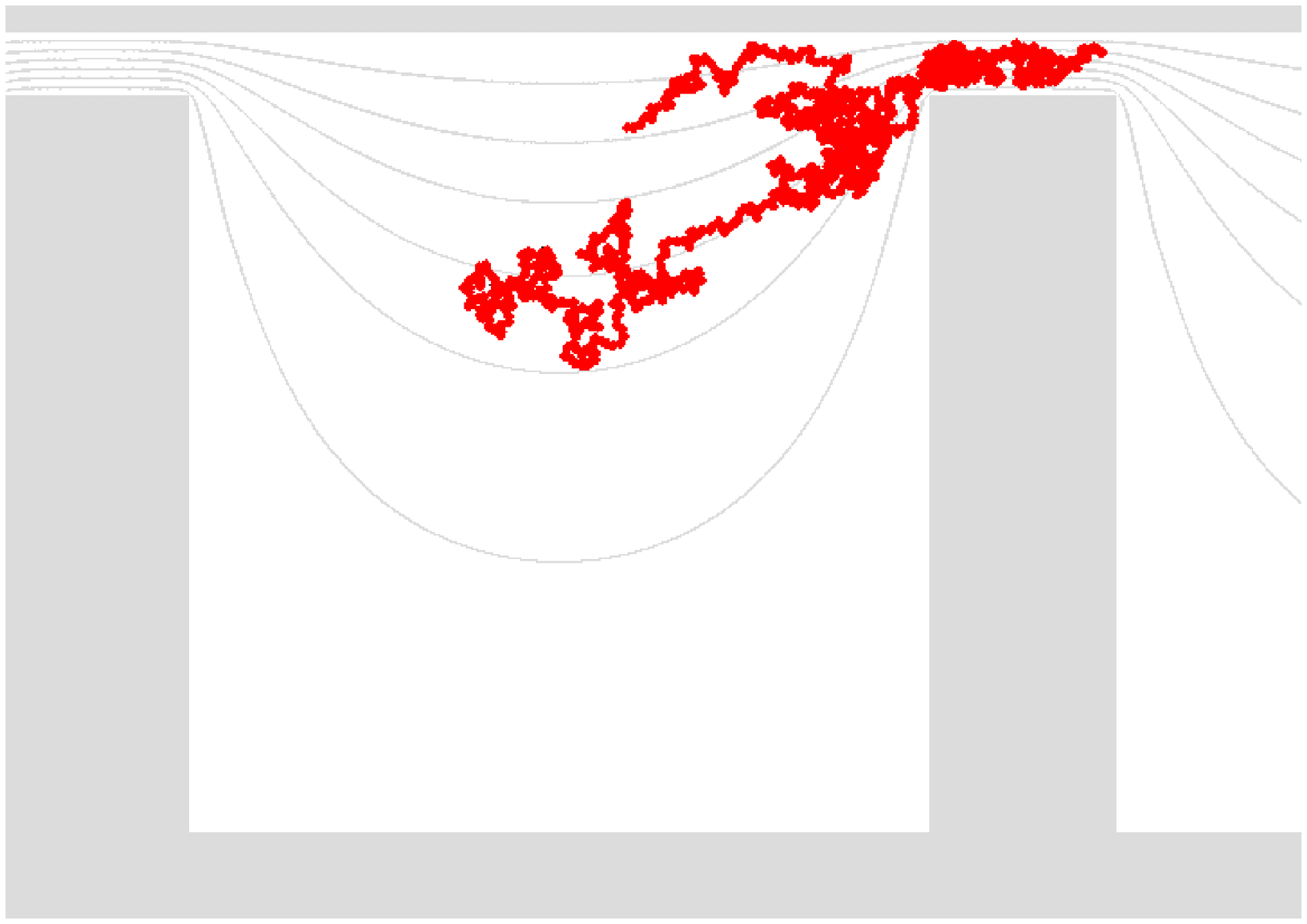}}
\CCsnapshot
\end{figure}

DNA is a charged polyelectrolyte with $2e^-$ per base pair, 
thus each bead carries a charge $q$ and is subject to an electric 
field \mbox{$\EE =  -\nabla \Phi$}. We assume that the charges do not 
interact with one another. This will be discussed further below.
The distribution of the electric potential $\Phi(\rr)$ in the 
channel was calculated numerically by solving the Laplace 
equation \mbox{($\Delta \Phi = 0$)} with von Neumann boundary 
conditions at the walls \mbox{($\nn \cdot \nabla \Phi = 0$},
where $\nn$ is the surface normal). 

The dynamical evolution of the system is described by a
Kramer's equations (Risken, 1989)
\begin{eqnarray}
\label{kramer}
\dot{\rr}_i & = & \vv_i \\
m \dot{\vv}_i & = & \ff_i - \zeta \vv_i + \eta_i,
\nonumber
\end{eqnarray}
where $\ff_i$ is the total force acting on bead $i$, and
$\vv_i$ its velocity, and $\zeta$ is a friction coefficient.
The random noise $\eta$ fulfills
\begin{eqnarray}
\langle \eta_{i \alpha}\rangle &=& 0 \\
\langle \eta_{i \alpha} (t) \eta_{j \beta}(t')\rangle 
&=& 2 k_B T \zeta
\delta_{ij} \: \delta_{\alpha \beta} \: \delta (t-t'),
\nonumber
\end{eqnarray}
with $\alpha, \beta = x,y$ or $z$. The random noise and 
the friction mimics the effect of the solvent surrounding
the DNA. The chains differ from standard Rouse chains (Doi, 1986)
by their excluded volume interactions and by the inertia $m$
of the beads. With the parameters of our model
($m=1. \zeta^2 \sigma^2/k_B T$, see the discussion further below),
the chains exhibit Rouse dynamics on length scales of the
order the gyration radius of the chain, $R_g$ (Streek, 2002).

Several important physical factors are neglected in the model.
First, electrostatic interactions within DNA chains are not taken
into account. This is justified by the fact that the Debye screening 
length in typical electrophoresis buffers is only a few $nm$, 
comparable to the diameter of the double helix.  
Second, we do not consider electro-osmotic flow, or flow in general. 
This reflects the experimental situation reported by Han \etal. 
Third, hydrodynamic interactions are disregarded. This is
partly justified by the theoretical observation that for 
polyelectrolytes dragged by an electric field, the hydrodynamic 
interactions should be screened over distances larger than the Debye 
length, since the counterions are dragged in the opposite 
direction (Viovy, 2000). Unfortunately, the argument is only strictly
valid as long as no non-electric forces are present (Long, 1996). 
In our case, the walls of the channels exert non-electrical forces 
which stop the polymers, but do not prevent the counterions 
in the Debye layer from moving. Thus the chains experience an 
additional trapping force from the friction of the counterions.
This effect is disregarded in our model. Furthermore, diffusion
is not treated correctly even in free solution. The diffusion
constant of Rouse chains scales as $1/N$ with the chain length
$N$. Including hydrodynamic interactions, one expects Zimm
scaling $1/R_g$, where the gyration radius $R_g$ scales like
\mbox{$R_g \propto N^{3/5}$} for self-avoiding chains.
Experimentally, the diffusion constant of DNA in typical 
buffer solutions is found to scale as $1/N^{0.672}$ (Stellwagen, 2003).

Unfortunately, a full treatment which accounts correctly both 
for the (dynamically varying) counterion distribution as well 
as the hydrodynamic interactions is not feasible with standard
computational resources. The simplifications of our model 
influence the results quantitatively, but do presumably not 
change them by orders of magnitude. The qualitative behavior 
should not be affected.

Keeping these caveats in mind, we can now proceed to establish
a quantitative connection with the experimental setup of
Han \etal. The natural units in our model are related to
the parameters $\sigma$, $\zeta$, $|q|$ (the charge per bead), 
and $T$ (the temperature). More specifically, 
the energy unit is $\epsilon=k_B T$, the length unit is $\sigma$,
the time unit is \mbox{$t_0 = \zeta \sigma^2/k_B T$},
and the electric field unit is $E_0 = k_B T/\sigma |q|$. 
Throughout the paper, all quantities shall be given in 
terms of these natural units. They shall now be related 
to real (SI) units.

Since the experiments are carried out at room temperature,
the energy unit is $\epsilon \equiv 300 k_B K = 4 \cdot 10^{-21} J$. 
The length unit is obtained from matching the persistence
length of the model chains, $l_p = 1.6 \sigma$ (Streek, 2002)
with that of DNA, $l_p = 45 nm$, yielding $\sigma \equiv 30 nm$. 
The persistence of the chain is also used to determine the
number of base pairs (bp) per bead. A DNA molecule contains 
approximately 150 bp on a stretch of length $l_p$. 
In our model, the average bond length between two beads is 
$0.847 \sigma$ (Streek, 2002), thus we have 1.9 beads per
persistence length, and one bead represents roughly 80 bp.
The elongation for a chain crossing with minimal energy is
$r_{\mbox{\tiny cross}} \approx 1.198 \sigma$, which corresponds
to an energy barrier $\Delta E_{\mbox{\tiny cross}} \approx 82 E_0$
for crossing. The Boltzmann-factor for this energy barrier turns 
out to be less than $10^{-35}$. To check the simulation program,
the bond length distribution was compared to the Boltzmann-factor 
and very good agreement was found. Furthermore, no bond ever 
exceeded a length of $1.1 \sigma$. Thus no chain crossing occured 
in our simulations. 

The time scale $t_0$ is calculated from the diffusion 
constant $D$. For Rouse chains of length $N$, $D$ is given by
\begin{equation}
D = \frac{k_B T}{N \zeta} = \frac{\sigma^2}{N t_0}.
\end{equation}
Experimentally, Stellwagen \etal~(2003) have recently reported
the relation
\begin{equation}
D = 7.73 \cdot 10^{-6} (\mbox{number of base pairs})^{-0.672} cm^2s^{-1}.
\end{equation}
Choosing as reference a chain of length 40 kbp ($N=500$),
we obtain $t_0 \equiv 2.9 \cdot 10^{-6} s$.

Finally, the unit of the electric field can be identified
from matching the mobility $\mu_0$ of free chains.
The theoretical value is $\mu_0 = |q|/\zeta = \sigma/t_0 E_0$.
In experiments, the mobility depends strongly on the
choice of the buffer. Unfortunately, Han \etal~do not
report explicit measurements of the free-chain mobility of DNA
in the buffer used in their experiments. In the microchannel,
they observe that the overall mobility saturates at high field 
strengths. The apparent maximum mobility $\mu_{max}$ results
from an average over a slow motion in the reduced electric 
field of the deep regions, and a fast motion in the enhanced 
electric field of the constrictions
(see Fig.~\ref{fig:snapshot}). For large periods $L\gg t_l, t_s$, 
the ratio of these two field strengths is simply the inverse 
of the thickness ratio $\alpha = t_l/t_s$. One can then
derive a relation between the true free chain mobility $\mu_0$ 
and the apparent mobility $\mu_{max}$:
\begin{equation}
\label{eq:muapp}
\mu_0 = \mu_{max} \: (x_l + x_s/\alpha) (x_l + x_s \alpha),
\end{equation}
where $x_l$ and $x_s$ are the relative lengths for the
deep regions and shallow constrictions, respectively ($x_l + x_s = 1$).
Han \etal~measure $\mu_{max} \approx 0.13 \cdot 10^{-8} m^2/Vs$ 
in an experimental setup with $x_l = x_s = 1/2$, 
$\alpha \approx 15$, and $L = 10-40 \mu m$ (Han, 1999).
Using Eq. (\ref{eq:muapp}), one can thus estimate
$\mu_0 = 4.3 \mu_{max} = 0.55 \cdot 10^{-8} m^2/Vs$.
This value of $\mu_0$ seems very low compared to
typical values in the literature. Stellwagen \etal~have 
measured DNA mobilities in Tris-acetate buffers 
at various concentrations and found values between 
$2-4 \cdot 10^{-8} m^2/Vs$ (Stellwagen, 2002).
In 40 mM Tris-acetate buffer, they obtain
$\mu_0 = 3.3 \cdot 10^{-4} cm^2/Vs$ (Stellwagen, 2003). 
Using the latter value as an order-of-magnitude estimate, 
we identify  $E_0 \sim 3. \cdot 10^3 V/cm$. However, the results
of Stellwagen \etal~can probably not be applied here, because
the mobility depends strongly on the buffer. The first estimate,
$\mu_0 = 0.55 \cdot 10^{-8} m^2/Vs$, leads to the
identification $E_0 \sim 2. \cdot 10^4 V/cm$. 

Han \etal~(1999, 2000) have separated DNA of lengths between 
5 and 160 kbp. Here we study chains of length $N \le 1000$, 
corresponding to $\le$ 80 kbp, which is comparable. 
The depth of the deep channel regions in the experiments
was $t_l = 1-3 \mu m$, which also compares well with the 
simulation, $80 \sigma \equiv 1.6 \mu m$. 
The depth of the constriction was $t_s = 100 nm \approx 2 l_p$
in the experiments. In our case, it is $5 \sigma \approx 3 l_p$, 
which is slightly larger, but still comparable. Since
our channels are wider, we expect that the trapping in
the simulations will be less pronounced than in reality.
The total length per trap (period) was 
$L = 4-40 \mu m$ in the experiments (mostly $4 \mu m$)
and $L = 100 \sigma \equiv 3 \mu m$ in the simulations. 

The average electric field strength in the experiments
was varied between 20 and 80 $V/cm$, which corresponds
to $\sim 0.001-0.004 E_0$ or $\sim 0.006-0.03 E_0$,
depending on the identification of $E_0$. In the simulations, 
we studied field strengths between $0.0025-0.04 E_0$. 
When comparing field strengths, we must keep in mind that the 
electric field in the channel is not homogeneous (see 
Fig.~\ref{fig:snapshot}).
In our geometry, the electric field in the constriction
is enhanced by a factor of 2.5. In the experimental 
setup, the enhancement factor is only $\sim 1.8$, due
to the fact that the length ratio between the shallow 
and deep regions is different (50:50 in the experiment,
20:80 in the simulations). The ratio of field strengths 
in the shallow and deep regions in our simulations
is roughly 4. In the experimental geometry, it is
$t_l/t_s \approx 10-30$ for large periods $L$ and
smaller otherwise. 

The remaining model parameter that has yet to be determined 
is the mass $m$ of a bead. We note that the actual value of the 
mass has no influence on the static properties of the chain 
({\em e.~g.}, the chain flexibility), nor on the diffusive part 
of the dynamics. It does, however, determine the relative importance 
of vibrational modes in the chain and other inertia effects. 
The latter can be characterized by the electrophoretic
relaxation time $\tau_e$, {\em i.~e.}, the characteristic 
decay time of the drift velocity of free flow DNA, if the electric
field is suddenly turned off. Typical values of $\tau_e$
are $\tau_e \approx 10^{-9}-10^{-12}s$ (Grossmann, 1992). 
In our model, the electrophoretic relaxation time is 
$\tau_e = m/\zeta$, and the correct value of the 
mass would be $m \sim 10^{-3}-10^{-6} \zeta t_0$. However, 
this is unfortunate from a computational point of view, 
because the mean velocity per bead, 
$\sqrt{\langle v^2 \rangle} = \sqrt{2 k_B T/m}$, becomes
large for such small bead masses, and the time step
has to be chosen short as a consequence. The simulation
becomes very inefficient. On the other hand, we are not interested
in inertia effects here, and we wish to study dynamical
effects on much longer time scales than $t_0$. Therefore, we chose 
to give the beads an unphysically high mass, $m = 1 \zeta t_0$, 
leading to an electrophoretic relaxation time 
$\tau_e = t_0 \sim 10^{-6}s$. On time scales $t_0$ and less, 
the dynamics will thus be unrealistic, but this does not affect the 
slow diffusive dynamics.

We close this section with a few technical remarks.
The dynamic equations (\ref{kramer}) were integrated
using the Verlet algorithm. The stochastic noise was
realized by picking in every time step a vector $\eta$ 
at random. D\"unweg and Paul (1991) have shown that 
the distribution of random numbers in such a procedure 
does not necessarily have to be Gaussian. Here, we
used a uniform distribution in a sphere. Since we
consider single chains only, no periodic boundaries
were necessary. With the mass $m = 1 \zeta t_0$, 
the time step could be chosen 0.01 $t_0$. 
Typical run lengths were between 4 and 20 million 
$t_0$ (5-25 seconds). The simulation jobs were
managed by the Condor Software Program (Condor),
which was developed by the Condor Team at the Computer
Science Department of the University of 
Wisconsin (Condor, 2003).

\section{Results and Discussion}
\label{sec:results}

Three examples of trajectories at the average field 
$E = 0.005 E_0$ are shown in Fig.~\ref{fig:trajectory}.
They reveal three qualitatively different modes of
migration. The trajectory of very short chains
($N=10$ corresponding to 800 bp) is dominated by
diffusion. The chains wander back and forth in the
trap, until they eventually escape into the next trap.
The movement of chains with intermediate chain length 
($N=100$ or 8 kbp) is much more directional, but still
irregular. They are often trapped at the entrance 
of constrictions. In contrast, long chains 
($N=1000$ or 80 kbp) travel smoothly. The first (leading) 
monomer is sometimes trapped, but this does not 
arrest the rest of the chain. Whereas the time 
spent in one box fluctuates strongly for shorter
chains, it is roughly constant for large chains.

\newcommand{\CCtrajectory}
{
\caption{
Trajectories of chains moving in an entropic trap 
array for three different chain lengths $N$ in the
presence of an average electric field $E = 0.005 E_0$. 
The middle solid line shows the position of the
center of mass. The (dashed) upper and lower lines show
the positions of the first and the last monomer.
In the case $N=10$, these three lines cannot be
distinguished from each other.
The dashed horizontal lines indicate the positions where 
constrictions begin. 
}
\label{fig:trajectory}
\bigskip
}
\begin{figure}[t]
\noindent
\includegraphics[width=0.48\textwidth]{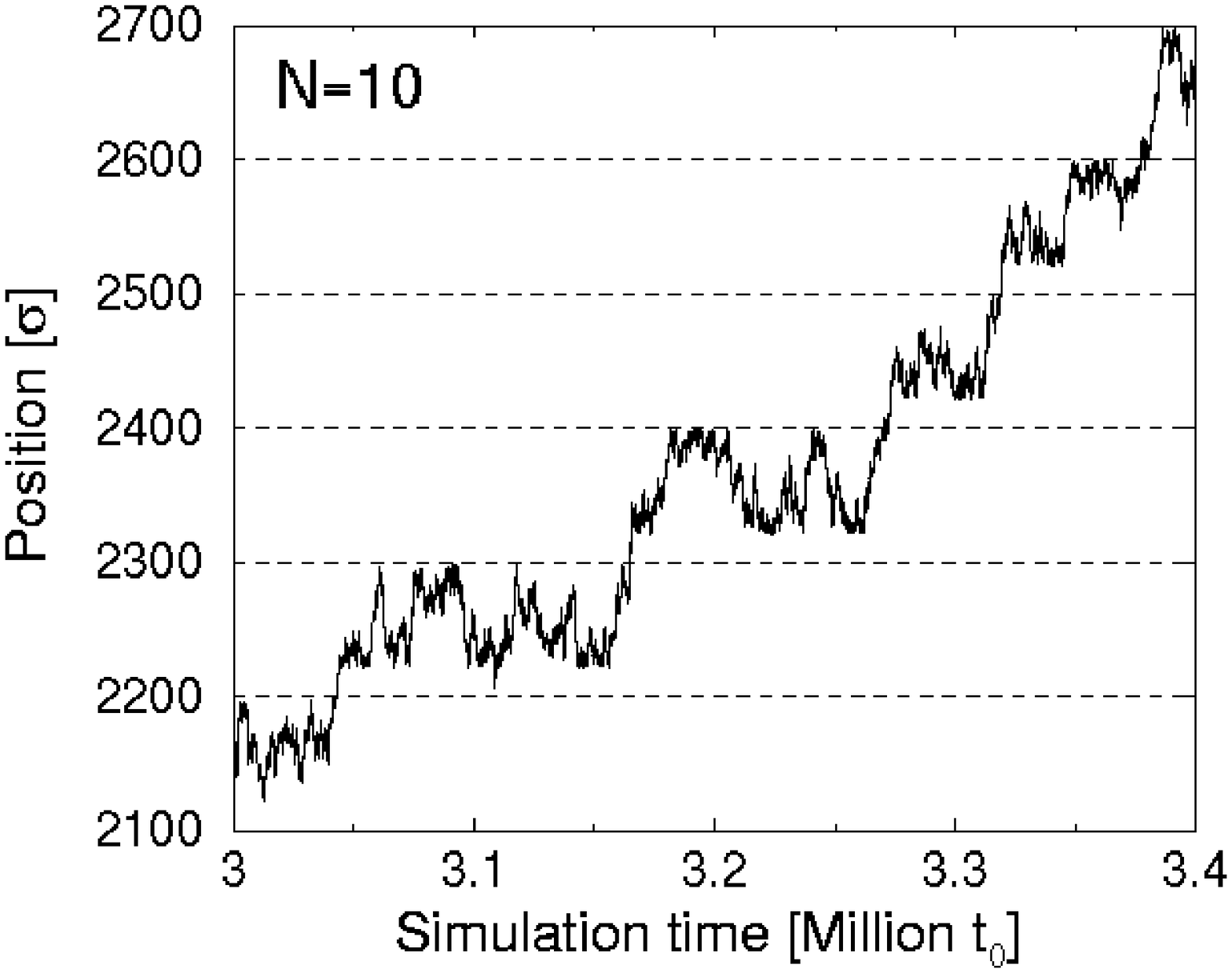}
\includegraphics[width=0.48\textwidth]{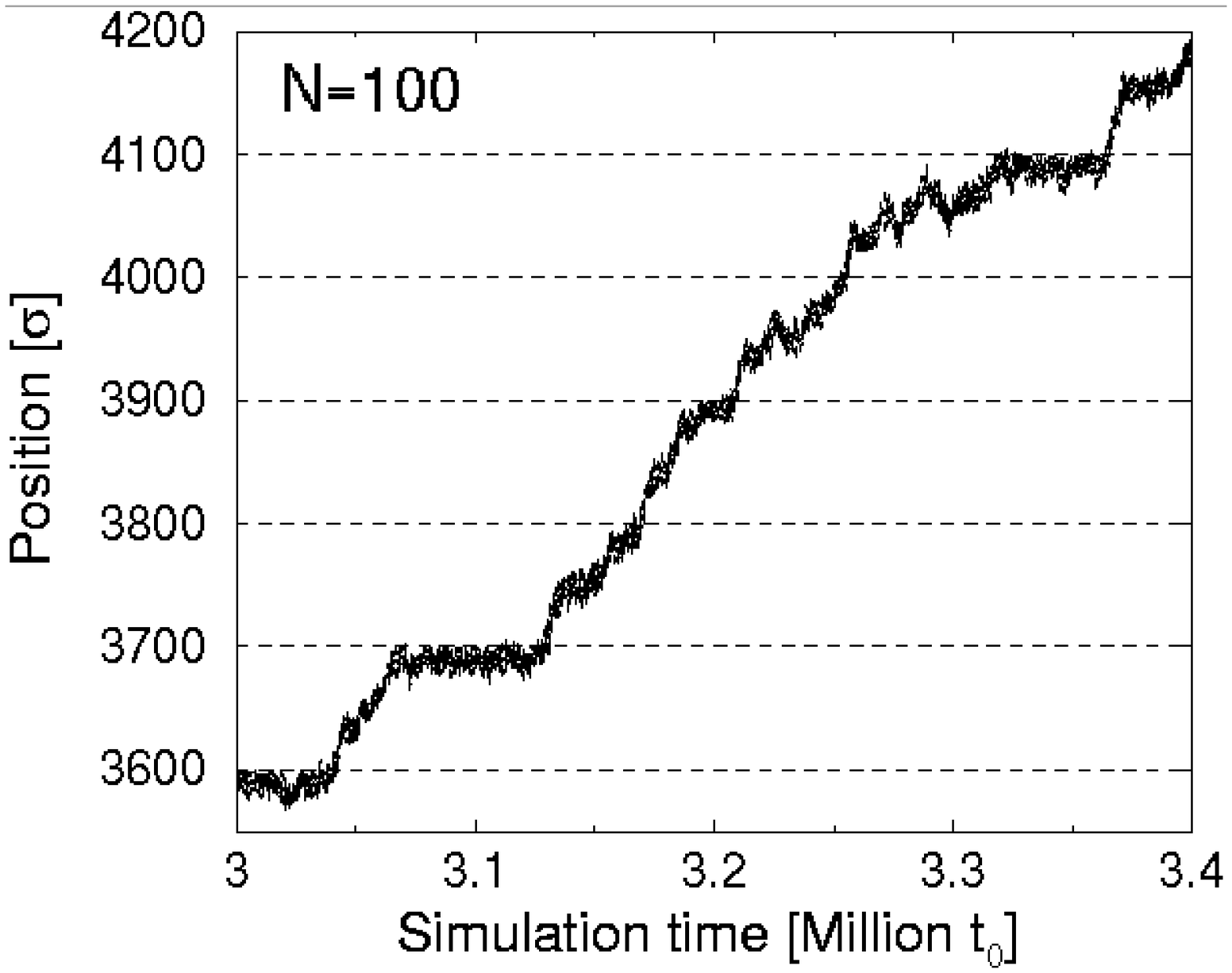}
\centerline{\includegraphics[width=0.48\textwidth]{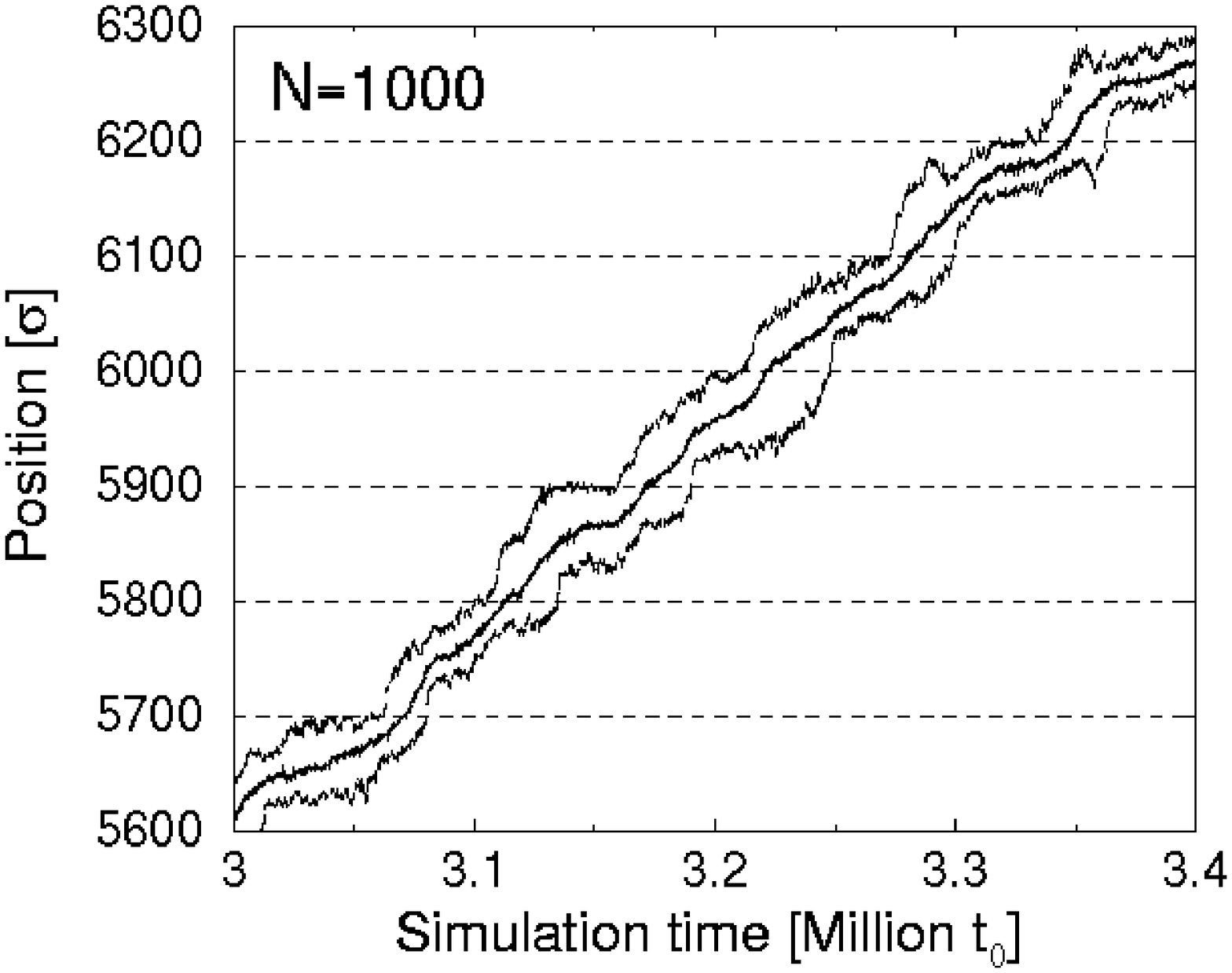}}
\CCtrajectory
\end{figure}

The behavior of intermediate chains ($N=100$) has similarity 
to that observed in the simulations of Tessier \etal~(2002). 
However, the trapping is much less pronounced in our case, 
and chains frequently pass from one box to another without being 
trapped at all. Trapping effects comparable to those reported 
by Tessier \etal~were only observed at the lowest field
strength, $E = 0.0025 E_0$. At that field value, chains of 
all lengths got trapped. As we shall see below, however, 
chain separation turned out to be not efficient for such 
small fields.

We have carried out simulations for seven different chain
lengths $N=$ 10, 20, 50, 100, 200, 500, and 1000, and
for five average field values, $E/E_0$ = 0.0025,
0.005, 0.01, 0.02, and 0.04. The resulting mobilities, 
determined as $\mu = \langle v \rangle/E$, are summarized
in Fig.~\ref{fig:mobility}. For all fields except the 
lowest, the mobility increases steadily with the chain 
length. At high chain lengths and high fields, it begins 
to saturate. The maximum value $\mu_{max}$ is only about 
half as large as the free chain mobility $\mu_0$, due to
the fact that the chains spend a disproportionate amount 
of time in the deep regions, where the local electric 
field is lower than average.

\newcommand{\CCmobility}
{
\caption{
Mobility $\mu$ in units of the mobility $\mu_0$ of a
free chain as a function of chain length $N$ for
different average electric fields $E$ in units
of $E_0$. The dashed lines show the prediction
of Eq.~(\ref{eq:model}) for the fields
$E/E_0 = 0.04, 0.02, 0.01, 0.005$ (from top to bottom).
}
\label{fig:mobility}
\bigskip
}
\begin{figure}[t]
\noindent
\centerline{\includegraphics[width=0.7\textwidth]{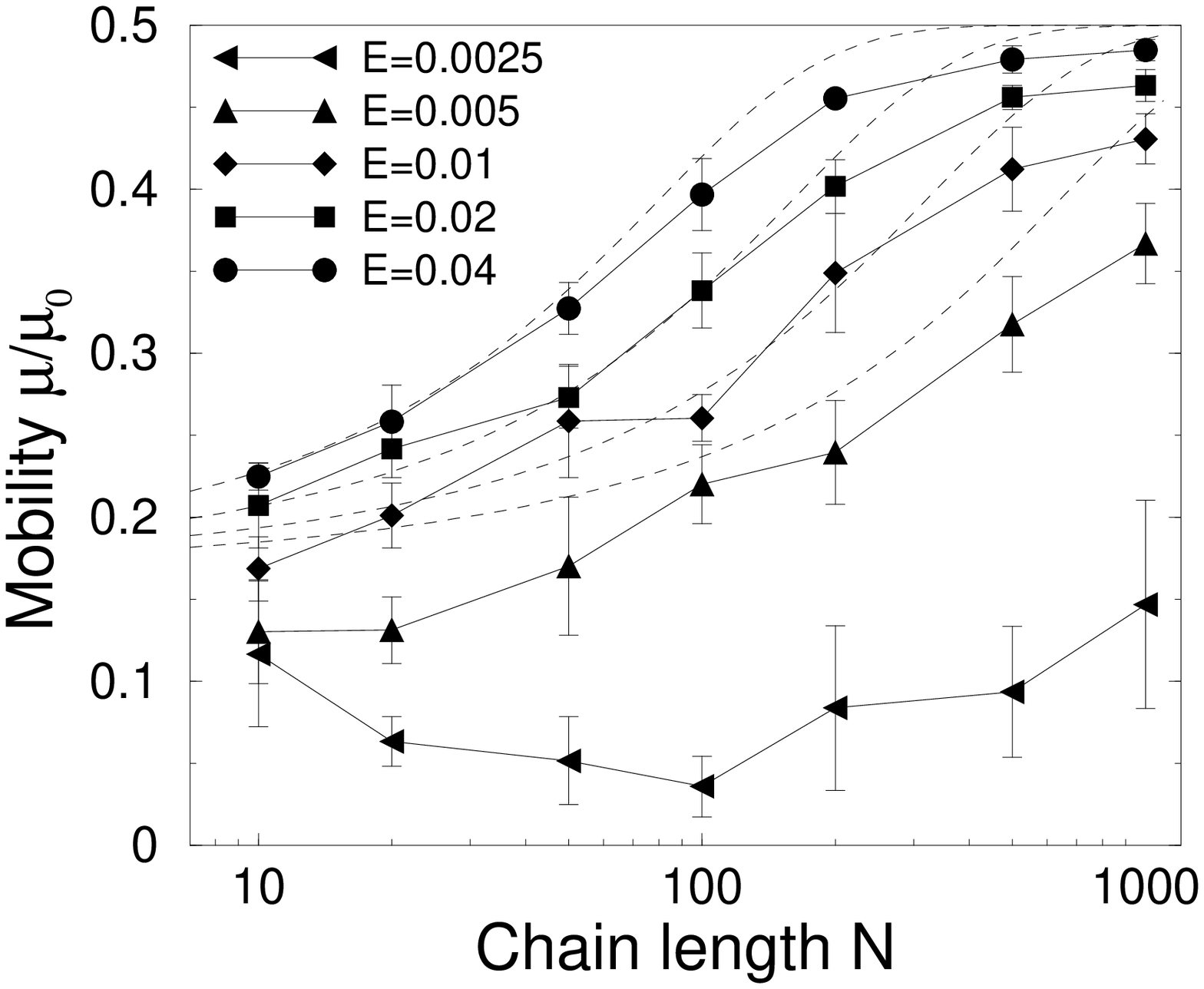}}
\CCmobility
\end{figure}

At the lowest field ($E=0.0025 E_0$), the mobility depends 
only slightly on the chain length and even decreases 
with chain length for small $N$. At such small fields,
backwards diffusion becomes important for short chains.
We have seen in Fig.~\ref{fig:trajectory} that short chains 
explore the whole trap. At $E = 0.0025 E_0$, they 
sometimes even travel backwards into the trap that they just 
left. The conformational entropic penalty for entering the 
constrictions is small for short chains. In the limit 
$E \to 0$, the inverse mobility is thus simply proportional
to the number of times the chain visits the entrance of the 
constriction. Since the latter scales like $N$ (the inverse 
diffusion constant $D^{-1}$), the mobility is then expected to 
decrease with chain length. In our system, this is observed at 
$E = 0.0025 E_0$ for chain lengths smaller than $N=100$.
For $N > 100$, the mobility increases again with chain
length. The resulting overall chain length dependence 
is small, and the chain separation is not efficient.

The quality of molecular separation systems is often 
characterized in terms of the theoretical plate number
\begin{equation}
\label{plate}
{\bf N}_{plate} = 16 \: (t_R/t_W)^2.
\end{equation}
where $t_R$ is the retention time, {\em i.~e.}, the total
time spent in the system, and $t_W$ the width of the peak
at the baseline. Fig.~\ref{fig:plates} shows the plate
number per trap for our system. In the interesting
regime, we have 10-100 plates per trap. At 
$\sim 10^5$ traps per meter, this corresponds to
theoretical plate numbers of $10^6-10^7$ plates/m,
which is quite good and in agreement with the results 
of Han \etal~(2000).

\newcommand{\CCplates}
{
\caption{
Theoretical plate number per trap as a function of
chain length for different electric fields $E$
as indicated (in units $E_0$).
}
\label{fig:plates}
\bigskip
}
\begin{figure}[t]
\noindent
\centerline{\includegraphics[width=0.7\textwidth]{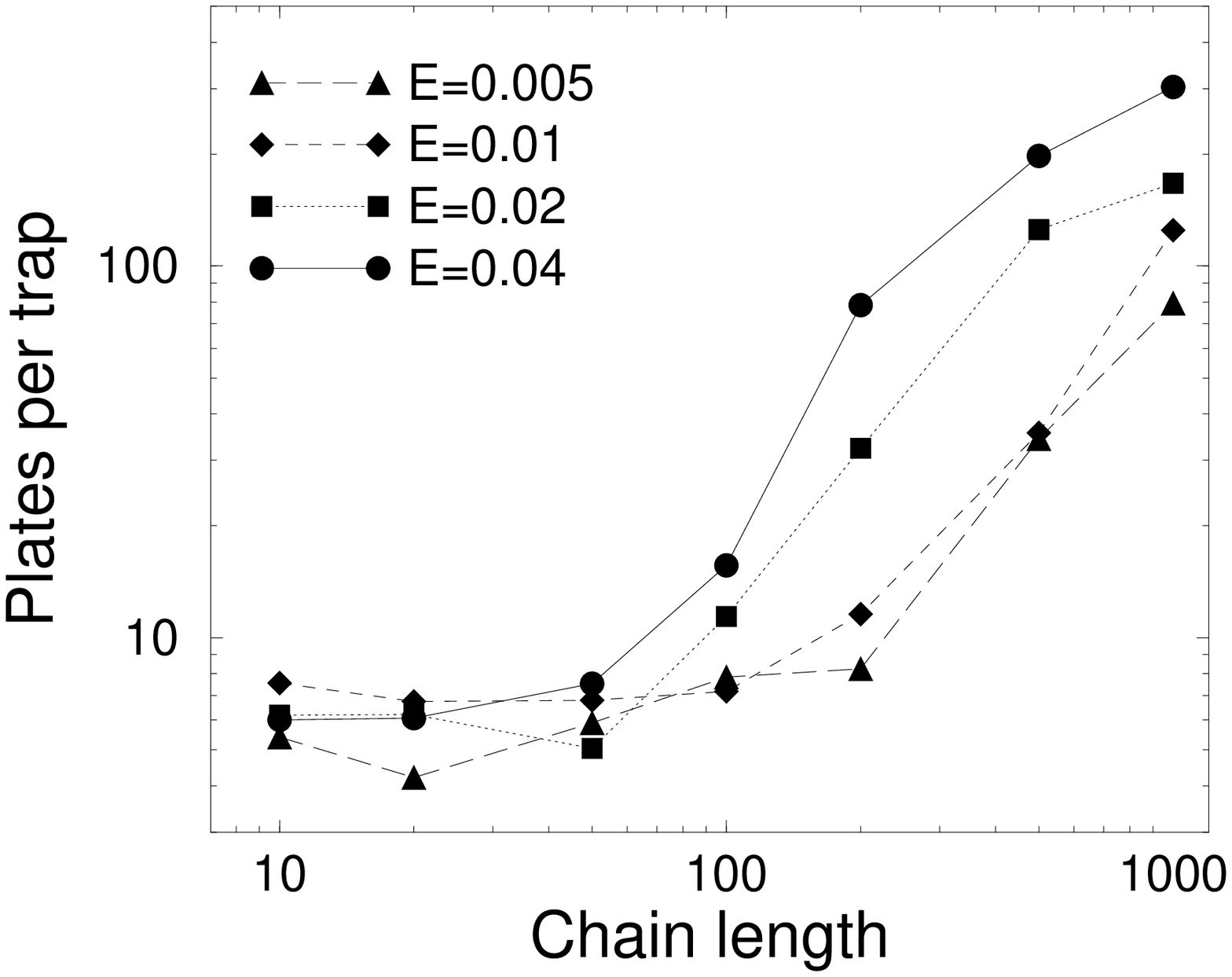}}
\CCplates
\end{figure}

We will now investigate the migration modes in more
detail. To this end, we have calculated histograms of the 
retention times spent in one trap. They were defined as the 
difference $t_{n+1}-t_n$ of the times $t_n$ when the 
first monomer of a chain first enters the deep region 
of the $n$th trap. Fig.~\ref{fig:histo} shows distributions 
of retention times for chains of different length $N$ 
in the field $E = 0.01 E_0$. 

\newcommand{\CChisto}
{
\caption{
Distribution of Retention times in one trap for
different chain lengths $N$ and electric field
$E = 0.01 E_0$. The thick solid line is a fit
to the initial exponential decay at chain length
$N=100$.
}
\label{fig:histo}
\bigskip
}
\begin{figure}[t]
\noindent
\centerline{\includegraphics[width=0.7\textwidth]{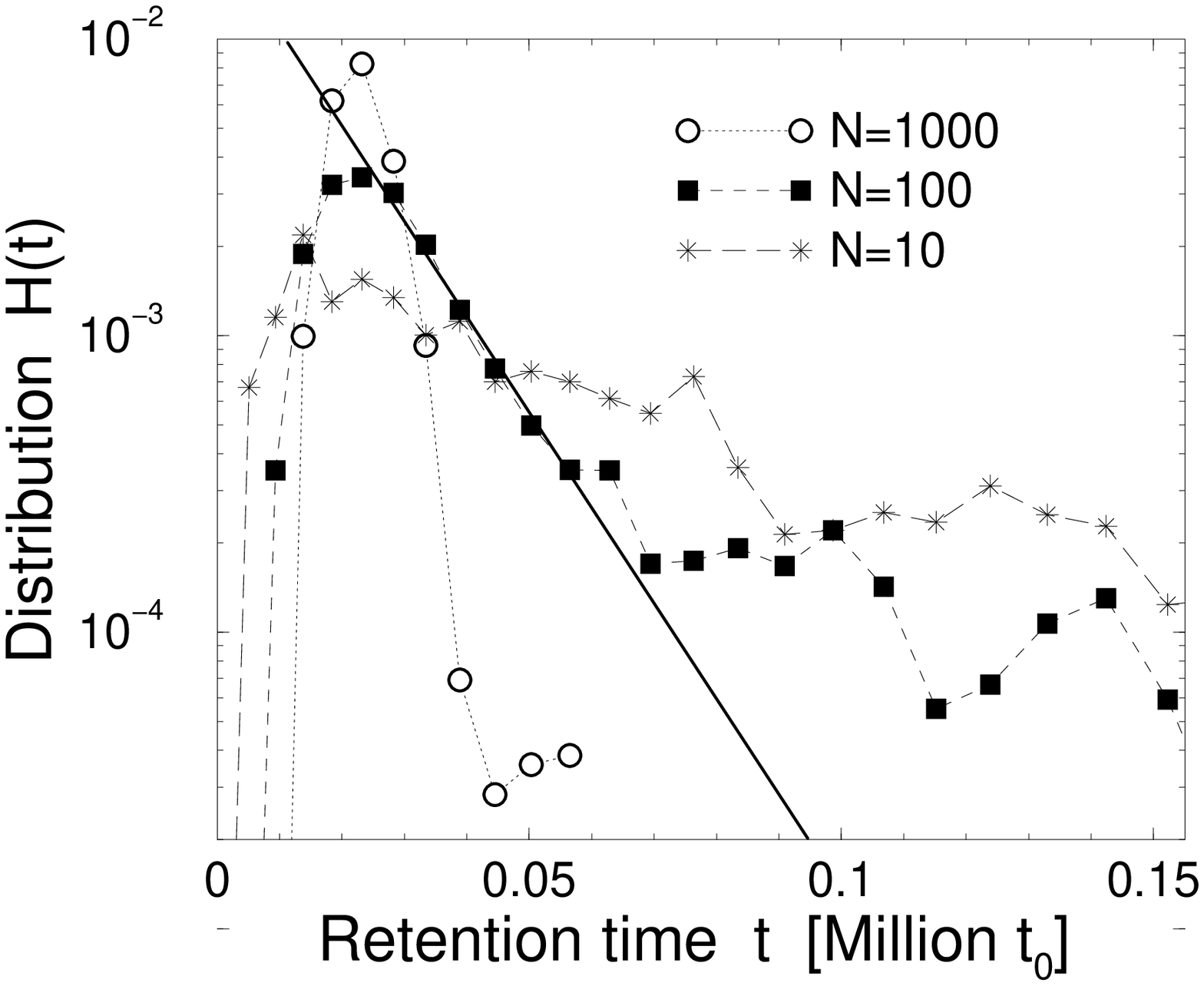}}
\CChisto
\end{figure}

The chains need a minimum time $\sim 1-1.5 \cdot 10^4 t_0$ 
to travel from one trap to another. After that ``dead''
period, the histogram rises rapidly and reaches a 
maximum at $t_{max} \approx 2. \cdot 10^4 t_0$.
>From a comparison of histograms for all our simulation
data (not shown), we deduce that the position of the maximum
$t_{max}$ of the distribution depends very little on the 
chain length and is strictly proportional to the inverse 
field $1/E$. The product $L/t_{max} E  = \mu_{max} \approx 0.5 \mu_0$ 
determines the maximum mobility $\mu_{max}$ in our system.

Beyond the maximum, the distribution decays rapidly 
for long chains ($N=1000$), and much more slowly for short 
chains ($N=10$), following an exponential behavior. This is 
consistent with the common picture, where the 
migration is determined by one single escape rate $1/\tau$. 
The histogram for intermediate chain length $N=100$, however, 
reveals that the situation is more complex in reality. The initial 
decay of the distribution can be fitted with one exponential, 
however, a long-time tail emerges at times beyond 
$t \sim 5 \cdot 10^4 t_0$. Thus the distribution of retention 
times at $N=100$ has {\em two} characteristic time scales 
$\tau_{\mbox{\tiny fast}}$ and $\tau_{\mbox{\tiny slow}}$. 

To some extent, this phenomenon is already apparent in the
trajectory of Fig.~\ref{fig:trajectory}. The figure suggests 
that there exist two qualitatively different ways how chains
pass from one trap to another: Either they travel relatively 
straight and unimpeded, or they get trapped and linger for 
some time at the border of the constriction. 

The trapping mechanism suggested by Han \etal~alone cannot 
explain these observations. Here we propose an additional 
trapping mechanism, which also slows down short chains 
and which presumably accounts to a large extent for the
chain length dependence of the mobility observed in our 
simulations. The idea is that chains get trapped at the 
side walls and corners of the deep boxes due to diffusion. 
The electric field lines lead the chains from the outlet of 
one constriction directly to the entrance of the next 
one. With a certain probability, the chain will 
therefore reach the next constriction without 
detours, and then get delayed there due to the mechanism 
suggested by Han \etal. This accounts for the 
fast time scale $\tau_{\mbox{\tiny fast}}$. On the other
hand, chains may also diffuse out of the main path.
They may access regions of the trap at the walls and in 
the corners where the electric field is very small. 
In that case, they are caught in a force-free region, 
and they can ``escape'' only by diffusion. 

We will now explore whether our data support this picture.
In the following analysis, only data for chain lengths
$N \ge 20$ and fields $E \ge 0.005 E_0$ were used.
In most of these systems two time scales $\tau$ were observed.
The value of the fast time scale $\tau_{\mbox{\tiny fast}}$ 
could be extracted by fitting an exponential 
\mbox{($A \exp(-t/\tau_{\mbox{\tiny fast}})$)} to the initial 
decay of the histogram $H(t)$. The determination of the
slow time scale was much more difficult, due to the poor 
statistics for the late time tails of the histograms.
We used two approaches: First we fitted the long 
range tail with an exponential function to obtain a rough 
estimate. Assuming $H(t) \propto e^{-t/\tau_{\mbox{\tiny slow}}}$, 
we calculated $\tau_{\mbox{\tiny slow}}$ via
\begin{equation}
\tau_{\mbox{\tiny slow}} = 
\int_{t_{\mbox{\tiny cut}}}^{\infty} dt \: 
(t - t_{\mbox{\tiny cut}})
\: H(t) \: \Big/
\int_{t_{\mbox{\tiny cut}}}^{\infty} dt \: H(t) 
\end{equation}
which is independent of $t_{\mbox{\tiny cut}}$. We used 
$t_{\mbox{\tiny cut}} = 500/E \: (t_0 E_0)$ to analyze the data. 
We checked whether the result deviated strongly from the previous
estimate, and whether it depended strongly on the
cutoff $t_{\mbox{\tiny cut}}$. If this was the case,
(because the data for $H(t)$ scattered strongly), the
result was discarded. The remaining values are compiled
in Fig.~\ref{fig:tau}, together with the data for
$\tau_{\mbox{\tiny fast}}$.

If our suspicion is correct that the fast process 
corresponds to the mechanism of Han \etal, then the rate 
$1/\tau_{\mbox{\tiny fast}}$ should be proportional to the 
amount of polymer in contact with the channel. Since the 
channel entrance is essentially one dimensional, the contact 
area should be proportional to the gyration radius $R_g$ of 
the chain.  Fig.~\ref{fig:tau} shows that the fast time 
scale indeed scales like $\tau_{\mbox{\tiny fast}} \propto
1/R_g \propto N^{-3/5}$. 

\newcommand{\CCtau}
{
\caption{
Characteristic length scales (rescaled) of the retention 
time distribution as a function of chain lengths $N$ 
for different electric fields $E$. The filled symbols 
correspond to the fast time scale $\tau_{\mbox{\tiny fast}}$,
the open symbols to the slow time scale
$\tau_{\mbox{\tiny slow}}$. The thick lines 
show for comparison power laws as indicated.
}
\label{fig:tau}
\bigskip
}
\begin{figure}[t]
\noindent
\centerline{\includegraphics[width=0.7\textwidth]{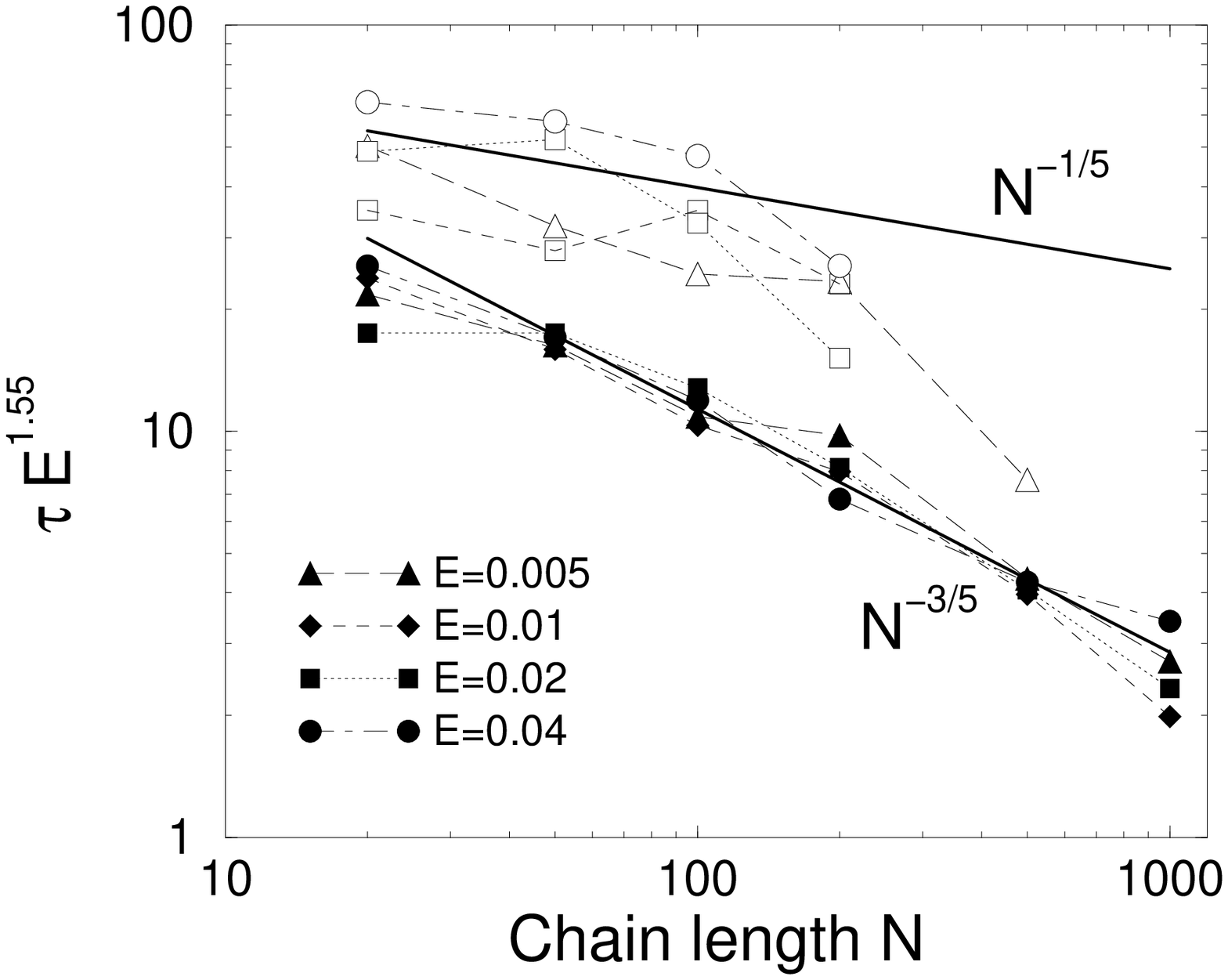}}
\CCtau
\end{figure}

The relation between 
$\tau_{\mbox{\tiny fast}}$ and the electric field 
strength $E$ is more complicated. According to 
Han \etal, the chains must overcome a free energy
barrier of height proportional to $1/E$ in order to
escape. On the other hand, the prefactor $\tau_0$ in 
Eq.~(\ref{eq:tau}) may also depend on $E$. The resulting
$E$-dependence can be rather complex. In the field range 
of our simulation, the resulting relation 
\mbox{$\tau_{\mbox{\tiny fast}}(E)$}
can be approximated by the empirical law 
\mbox{$\tau_{\mbox{\tiny fast}}\propto E^{-1.55}$}.

Like the fast time scale \mbox{$\tau_{\mbox{\tiny fast}}$},
the slow time scale \mbox{$\tau_{\mbox{\tiny slow}}$}
also decreases with the chain length, but the dependence
here is very weak. Unfortunately, the quality of the data 
was not sufficient for a quantitative analysis. If our 
interpretation is correct, then \mbox{$\tau_{\mbox{\tiny slow}}$}
characterizes an escape from a low-force region.
We note that extended chains in a corner experience
a net electric force towards the wall, even though
the field lines of course never enter the wall 
(see Fig.~\ref{fig:trap}). In order to leave the
corner, the chain must either move against the
force, or change its shape. In both cases, it
has to overcome a free energy barrier before it
can get rescued from the field lines. A simple
Ansatz would predict the escape probability 
to be proportional to the diffusion constant 
$D \propto 1/N$, and to the area covered by the 
chain, $R_g^2 \propto N^{6/5}$, or the total chain 
length $N$. This would yield a very weak net chain length 
dependence $\tau_{\mbox{\tiny slow}} \sim N^{-1/5}$ or 
$\tau_{\mbox{\tiny slow}} \sim N^0$, which is
consistent with the observed behavior (Fig.~\ref{fig:tau}).

\newcommand{\CCtrap}
{
\caption{
Schematic cartoon of a chain caught at the corner of the trap in 
the field-free region.  The chain experiences a net force towards the wall.
}
\label{fig:trap}
\bigskip
}
\begin{figure}[t]
\noindent
\centerline{\includegraphics[width=50mm]{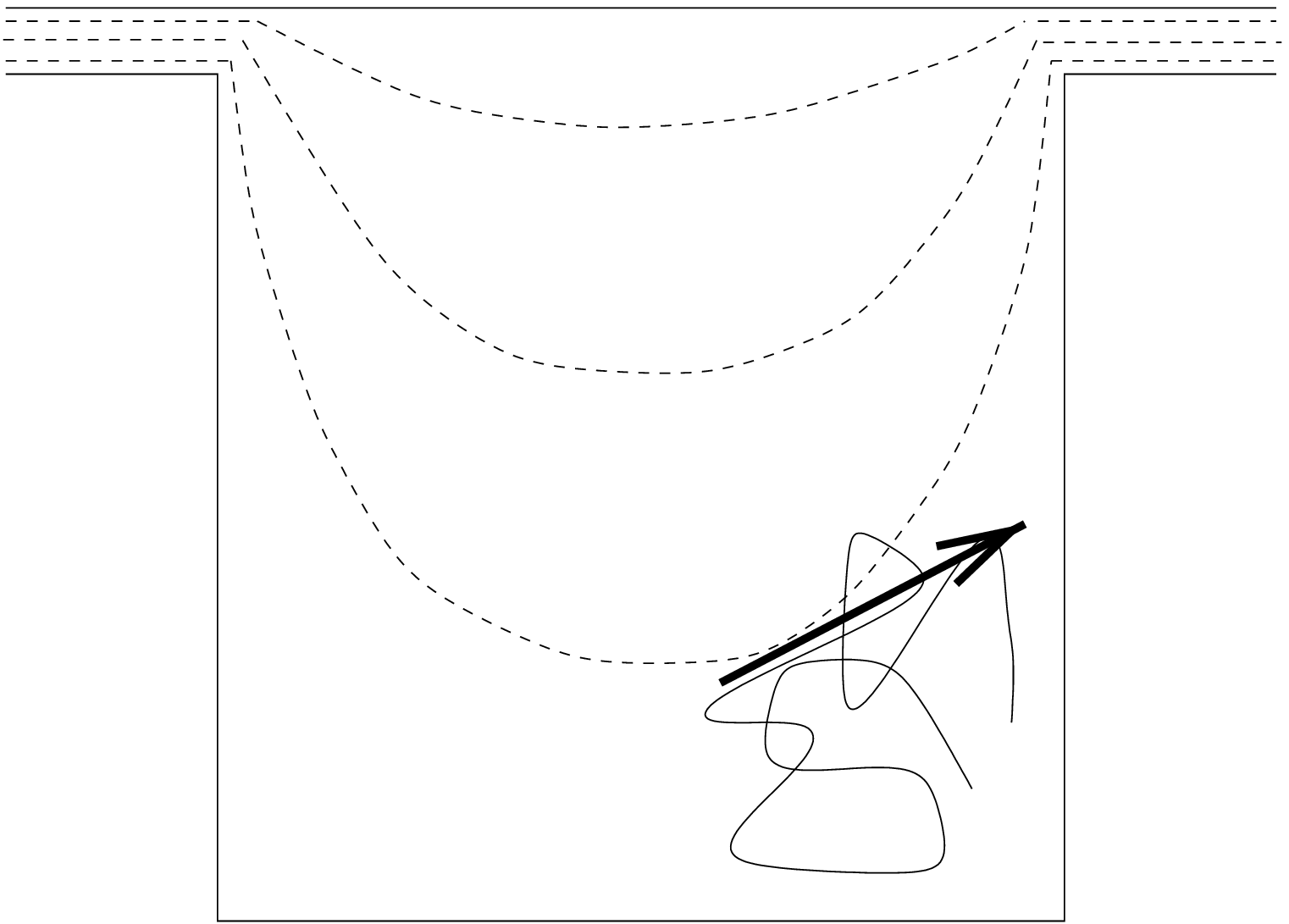}}
\CCtrap
\end{figure}

Thus the slow time scale \mbox{$\tau_{\mbox{\tiny slow}}$}
itself does not contribute much to the chain length
dependence of the mobility. The main effect comes
from the fact that the {\em relative number} of 
chains caught in the field-free  region
depends on the chain length. The travel time
from one channel to the next for undeflected chains
is slightly smaller than $t_{max} = L/E \mu_{max}$.
We assume that chains have to diffuse at least over
a distance $z_0$ into the field-free region (direction $-z$, see
Fig.~\ref{fig:device}) in order to get caught. The 
distribution along the $z$-direction after a time 
$t_{max}$ will be approximately Gaussian: 
$N(z) \propto e^{-z^2/6Dt}$. The probability of being caught is thus
\begin{equation}
\label{eq:erf}
P = \int_{z_0}^\infty N(z) = 
\erfc (z_0/\sqrt{6 D t_{max}})
= \erfc (\alpha \sqrt{N E}),
\end{equation}
where $\erfc(y) = 2/\sqrt{\pi} \: \int_y^{\infty} dx \exp(-x^2)$
is the complementary error function.

\newcommand{\CCpcum}
{
\caption{
Probability $P_0(t)$ that a chain is still in the trap
at the time $t_{\mbox{\tiny cut}}= 350/E \:(t_0 E_0)$,
vs. $N E$ in units of $E_0$. The solid line represents
a fit to Eq.~(\ref{eq:erf}) with $\alpha = 0.59$.
}
\label{fig:pcum}
\bigskip
}
\begin{figure}[t]
\noindent
\centerline{\includegraphics[width=0.7\textwidth]{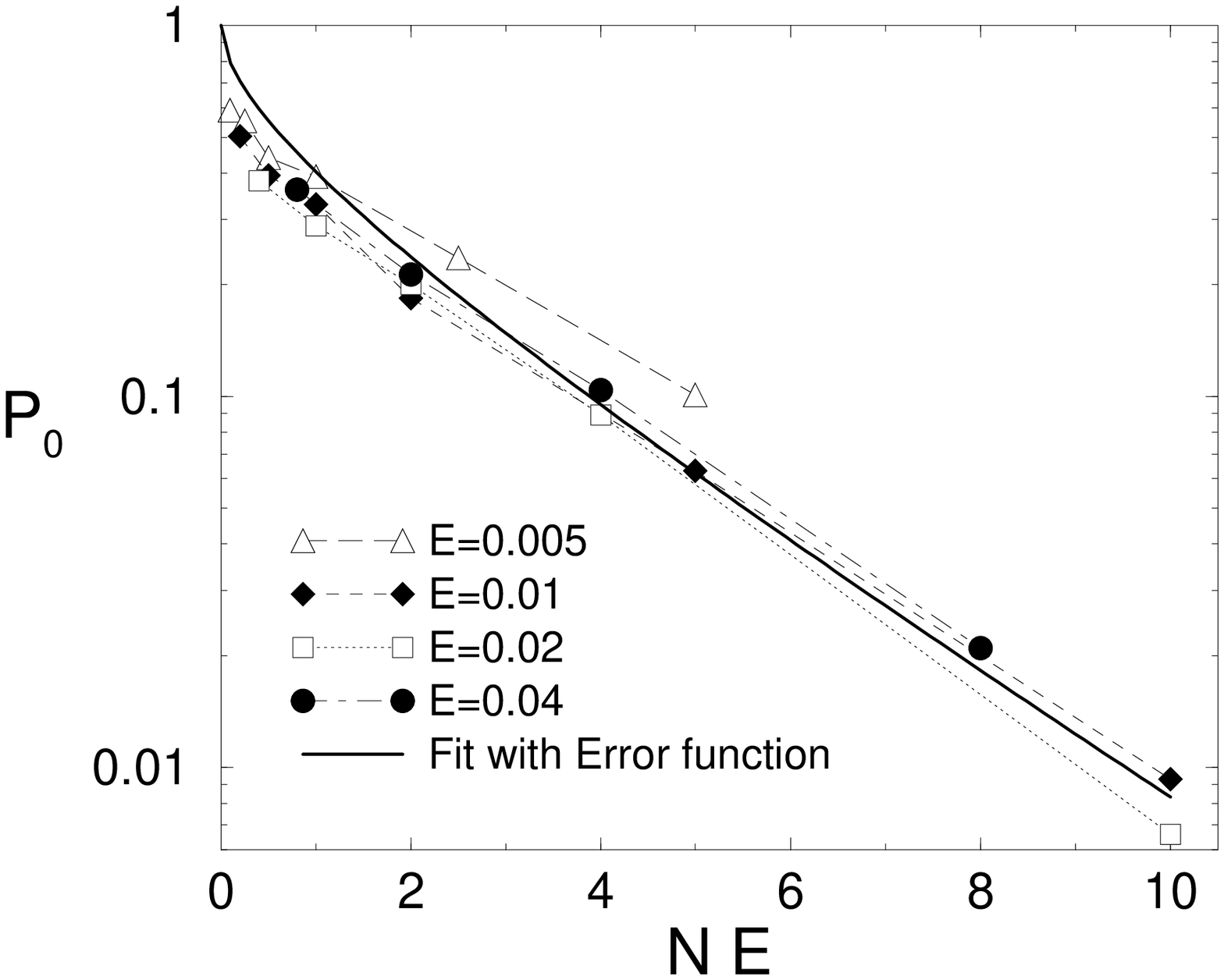}}
\CCpcum
\end{figure}

We can test this prediction under the assumption that 
the two time scales $\tau_{\mbox{\tiny fast}}$ and
$\tau_{\mbox{\tiny slow}}$ are sufficiently far apart
that they can be separated. In that case, one
can choose a time $t_{\mbox{\tiny cut}}$ such that almost 
all undeflected chains have left the trap, while almost 
all deflected chains are still in the field-free region. 
$P$ can then be approximated by the
relative number $P_0$ of chains left in the field-free region
at the time $t_{\mbox{\tiny cut}}$. Fig.~\ref{fig:pcum} 
shows the result of such an estimate. The data collapse 
reasonably well for different $N, E$ and can be fitted with 
Eq.~(\ref{eq:erf}), with the fit parameter $\alpha = 0.59$. 
Inserting $\mu_{max} = 0.5 \mu_0$ and $L = 100 \sigma$, 
one obtains $z_0 \sim 20 \sigma$. Thus chains get
caught if they sidetrack by more than
$\sim 20 \sigma$ from their main path, which is determined 
by the field lines.

These considerations establish the existence of a new 
mechanism which produces chain length dependent
mobility. To assess the relative importance of the
new mechanism, we compare the real mobility data, 
Fig.~(\ref{fig:mobility}), with a very simple model. 
Chains either travel straight across the trap, or
get sidetracked into the field-free region.
Traveling across the trap takes at least the time
$t_{max} = L E/\mu_{max}$. The chains caught in the
field-free region spend
an additional time $\tau_{\mbox{\tiny slow}}$ in
the trap. We make the simplification that 
$E \tau_{\mbox{\tiny slow}}$ is independent of 
the chain length and electric field and given by
the number $E \tau_{\mbox{\tiny slow}} = 400 \: t_0 E_0$,
which is roughly the value at chain length $\sim 100$
and field strength $\sim 0.01-0.02 \: E_0$.
The relative number of chains caught in the deep region
$P$ is calculated according
to Eq.~(\ref{eq:erf}), with $\alpha = 0.59$
(taken from Fig.~\ref{fig:pcum}). The resulting
mobility is

\begin{equation}
\label{eq:model}
\frac{\mu}{\mu_0} =
\left[ 
\frac{\mu_0}{\mu_{max}} + \frac{\tau_{\mbox{\tiny slow}} E}{L} \:
\erfc (\alpha \sqrt{N E}) \: 
\right]^{-1}.
\end{equation}

This prediction is compared with the actual data in
Fig.~\ref{fig:mobility} (dashed lines). Despite the
simplicity of Eq.~(\ref{eq:model}), the agreement
is remarkably good. 

\section{Summary and Outlook}
\label{sec:summary}

To summarize, we have presented the first off-lattice
Brownian dynamics simulation of DNA migration in an 
entropic trap array. We reproduce the experimentally
observed phenomenon that the mobility increases
with the chain length. This result can be traced
back to two distinct mechanisms. The first mechanism
has already been discussed by Han \etal: Chains get
delayed at the narrow channels connecting the traps.
They escape through the channels with a probability
which is proportional to the radius of gyration of
the chain and thus scales as $N^{3/5}$. However, we
found that this effect accounts only in part for the
total chain length dependence. In the second 
mechanism, the chains are trapped with a certain
probability at the side and in the corner of the box.
The characteristic time for escaping such a configuration 
is very long. The trapping probability increases with 
the diffusion constant, which is in turn inversely
proportional to the chain length. As a result, the
mobility increases with the chain length.

To our best knowledge, this mechanism has not yet 
been described in the literature. It becomes
relevant when the period $L$ of the structure
is small. Indeed, Han \etal~have studied structures
with periods ranging from $4$ to $40$ $\mu m$
(Han, 1999), but they reported separation by length 
only for their smallest structure with $L = 4 \mu m$, 
in a system with dimensions comparable to those
studied here.

We have observed a number of other phenomena, 
which we shall not describe in detail here. 
In contrast to Chen \etal~(Chen, 2003), we have
considered truly non-equilibrium systems.
Subsequent escapes cannot necessarily be considered 
as independent events. The longest chains $N=1000$
do not recover the equilibrium coil structure
in the middle of the trap, but they remain 
stretched in the $x$ direction. Moreover, our
data seem to provide evidence that chains even 
retain some memory of the previous escape process.
At high fields, successive escape times seem to
be correlated. Unfortunately, the statistical
quality of the data is not good enough to allow 
for a more thorough analysis. 

The situation becomes even more complicated when
the shallow channel is made wider. Whereas in the 
cases presented here, long chains migrated faster 
than short chains, we have observed the inverse
effect in microfluidic system with wider channels
(Duong, 2003). The effect as well as other, even 
more unexpected phenomena, can be reproduced in 
simulations. These phenomena will be presented in a
forthcoming publication (Streek, 2004).

\section*{Acknowledgments}

We thank Ralf Eichhorn for critically reading the
manuscript. This work was funded by the German Science 
Foundation (SFB 613, Teilprojekt D2). 

\section*{References}

\begin{list}{}{
  \setlength{\itemindent}{-\leftmargin}
  \setlength{\itemsep}{0.5 \baselineskip}
}

\item
  Bader, J.~S., Hammond, R.~W., Henck, S.~A., Deem, M.~W.,
  McDermott, G.~A., Bustillo, J.~M., Simpson, J.~W., Mulhern, G.~T.,
  Rothberg, J.~M., 1999.
  DNA transport by a micromachined Brownian ratchet device.
  PNAS 96, 13165-13169.

\item
  Chen, Z., Escobedo, F. A., 2003.
  Simulation of chain-length partitioning in a microfabricated
  channel via entropic trapping.
  Mol. Sim. 29, 417-425.

\item
  The Condor team, 2003. 
  software package from www.cs.wisc.edu/condor/.

\item
  Deutsch, J.~M., 1987.
  Dynamics of pulsed-field electrophoresis.
  Phys. Rev. Lett. 59, 1255-1258,

\item
  Deutsch, J.~M., 1988.
  Theoretical studies of DNA during gel electrophoresis.
  Science 240, 922-924.

\item
  Doi, M., Edwards, S.~H., 1986.
  The theory of polymer dynamics.
  Clarendon press, Oxford.

\item
  D\"unweg, B., Paul, W., 1991.
  Brownian dynamics simulation without Gaussian random numbers.
  Int. J. Mod. Phys. C 2, 817-827.

\item
  Duong, T. T., Kim G., Ros, R., Streek, M., Schmid, F.,
  Brugger, J., Anselmetti, D., Ros, A., 2003.
  Size-dependent free solution DNA electrophoresis in structured
  microfluidic systems. 
  Microelectronic Engineering 67-68, 905-912. 

\item 
  Grossmann, P.~D., Colburn, J.~C., 1992.
  Capillary Electrophoresis, Theory and Practice:
  Free Solution Capillary Electrophoresis.
  Acad. Press, San Diego.

\item
  Hammond, R.~W., Bader, J.~S., Henck, S.~A., Deem, M.~W.,
  McDermott, G.~A., Bustillo, J.~M., Rothberg, J.~M., 2000.
  Differential transport of DNA by a rectified Brownian motion
  device. 
  Electrophoresis 21, 74-80.

\item
  Han, J., Turner, S.~W., Craighead, H.~G.. 1999.
  Entropic trapping and escape of long DNA molecules at
  submicron size constriction.
  Phys. Rev. Lett. 83, 1688-1691;
  Erratum. 2001. Phys. Rev. Lett. 86, 1394.

\item
   Han., J., Craighead, H.~G., 2000.
   Separation of long DNA molecules in a microfabricated 
   entropic trap array. 
   Science 288, 1026-1029.

\item
  Han., J., Craighead, H.~G., 2002.
  Characterization and optimization of an entropic trap
  for DNA separation.
  Anal. Chem. 74, 394-401.

\item
  Long, D., Viovy, J.-L., Ajdari, A., 1996.
  Simultaneous action of electric fields and nonelectric forces
  on a polyelectrolye: motion and deformation.
  Phys. Rev. Lett. 76, 3858-3861.

\item
  Matsumoto, M., Doi., M., 1994.
  Brownian dynamics simulation of DNA gel electrophoresis
  Mol. Sim. 12, 219-226.

\item
  Noguchi, H., Takasu, M., 2001.
  Dynamics of DNA in entangled polymer solutions: An anisotropic
  friction model.
  J. Chem. Phys. 114, 7260-7266.

\item
  Risken, H. 1989. 
  The Fokker-Planck equation: Methods of solution and applications.
  Springer Verlag, Berlin.

\item
  Stellwagen, E., Stellwagen, N.~C., 2002.
  The free solution mobility of DNA in Tris-acetate-EDTA
  buffers of different concentration, with and without added NaCl.
  Electrophoresis 23, 1935-1941.

\item
  Stellwagen, E., Lu, Y., Stellwagen, N.~C., 2003.
  Unified description of electrophoresis and diffusion 
  for DNA and other polyions.
  Biochemistry 42, 11745-11750.

\item
  Streek, M. 2002.
  Migration of DNA on structured surfaces in an external field.
  Diploma thesis, Universit\"at Bielefeld.

\item
  Tessier, F., Labrie, J., Slater, G.~W., 2002.
  Electrophoretic separation of long polyelectrolytes
  in submolecular-size constrictions: A Monte Carlo study.
  Macromolecules 35, 4791-4800.

\item
  Viovy, J.-L., 2000. 
  Electrophoresis of DNA and other polyelectrolytes: Physical mechanisms. 
  Rev. Mod. Phys. 72, 813-872.

\end{list}

\end{document}